\documentclass[letter]{aa}  

\usepackage{graphicx}
\usepackage{txfonts}
\usepackage{lscape}
\usepackage{booktabs}
\usepackage{threeparttablex}
\usepackage{wasysym}
\usepackage{arydshln}
\usepackage{mathrsfs}
\usepackage{amssymb}
\usepackage[bookmarks=false,colorlinks=true,linkcolor=black,citecolor=cyan,filecolor=black,urlcolor=cyan]{hyperref}
\usepackage{array}

\newcolumntype{C}[1]{>{\centering\arraybackslash}p{#1}}

\begin{document} 
	
	
	\title{Cosmic hide and seek: the volumetric rate of X-ray quasi-periodic eruptions}
	
	\author{R. Arcodia\thanks{NASA Einstein fellow}
		\inst{1,2},
		A. Merloni
		\inst{2},
        J. Buchner
		\inst{2},
        P. Baldini
		\inst{2},
        G. Ponti
		\inst{3,2},
        A. Rau
		\inst{2},
        Z. Liu
		\inst{2},
        K. Nandra
		\inst{2},
        M. Salvato
		\inst{2}
	}
	\institute{MIT Kavli Institute for Astrophysics and Space Research, 70 Vassar Street, Cambridge, MA 02139, USA\\
		\email{rarcodia@mit.edu}
		\and
		Max-Planck-Institut f\"ur extraterrestrische Physik (MPE), Gie{\ss}enbachstra{\ss}e 1, 85748 Garching bei M\"unchen, Germany
        \and
        INAF-Osservatorio Astronomico di Brera, Via E. Bianchi 46, I-23807 Merate (LC), Italy
	}
	
	\date{Received ; accepted }
	
	
	\abstract{Multi-wavelength extragalactic nuclear transients, particularly those detectable as multi-messengers, are among the primary drivers for the next-generation observatories. X-ray quasi-periodic eruptions (QPEs) are the most recent and perhaps most peculiar addition to this group. Here, we report a first estimate of the volumetric rate of QPEs based on the first four discoveries with the eROSITA X-ray telescope onboard the Spectrum Roentgen Gamma observatory. Under the assumption, supported by a suite of simulated light curves, that these four sources sample the intrinsic population somewhat homogeneously, we correct for their detection efficiency and compute a QPE abundance of $\mathscr{R}_{{\rm vol}} = 0.60_{-0.43}^{+4.73} \times 10^{-6}\,$Mpc$^{-3}$ above an intrinsic average $\log L_{\rm 0.5-2.0\,keV}^{\rm peak} > 41.7$. Since the exact lifetime of QPEs ($\tau_{\rm life}$) is currently not better defined than between a few years or few decades, we convert this to a formation rate of $\mathscr{R}_{\rm vol}/\tau_{\rm life}\approx 0.6 \times 10^{-7} (\tau_{\rm life}/10\,\mathrm{y})^{-1}$\,Mpc$^{-3}$\,year$^{-1}$. As a comparison, this value is a factor $\sim10\,\tau_{\rm life}$ times smaller than the formation rate of tidal disruption events. The origin of QPEs is still debated, although lately most models suggest that they are the electromagnetic counterpart of extreme mass ratio inspirals (EMRIs). In this scenario, the QPE rate would thus be the first-ever constraint (i.e. a lower limit) to the EMRI rate from observations alone. Future discoveries of QPEs and advances in their theoretical modeling will consolidate or rule out their use for constraining the number of EMRIs detectable by the {\it LISA} mission.
    }
	
	
	\titlerunning{temp title}
	\authorrunning{R. Arcodia et al.} 
	\maketitle	

    \defcitealias{Miniutti+2019:qpe1}{M19}
	\defcitealias{Giustini+2020:qpe2}{G20}
	\defcitealias{Arcodia+2021:eroqpes}{A21}
	\defcitealias{Arcodia+2024:qpe34}{A24}
	
	\section{Introduction}
    \label{sec:intro}

    The interest in multi-wavelength extragalactic nuclear transients has recently increased with the advent of sensitive wide-area surveys, which started discovering them in significant numbers. Particular attention is being dedicated to transients that are detectable in a multi-messenger fashion, via electromagnetic radiation and gravitational waves (GW) and/or neutrino detectors. The main protagonists are Tidal Disruption Events (TDEs) with the recent addition of quasi-periodic eruptions (QPEs). For TDEs, high-cadence optical surveys, such as the Zwicky Transient Facility, took a major step forward in detection numbers \citep{VVelzen2021:ztf,Yao+2023:TDEs} after the first discoveries by ROSAT \citep{Bade:tde,Grupe+1999:tde}, although X-ray discoveries are catching up (\citealp{Sazonov+2021:erotde}; Grotova et al., in prep.) thanks to the extended ROentgen Survey with an Imaging Telescope Array \citep[eROSITA;][]{Predehl+2021:eROSITA} aboard the Spectrum-Roentgen-Gamma observatory \citep[\emph{SRG};][]{Sunyaev+2021:SRG}. For QPEs, the field is still in its early days, with the first discovery reported in 2019 \citep[][hereafter \citetalias{Miniutti+2019:qpe1}]{Miniutti+2019:qpe1} and only a handful of secure QPE sources known to date (\citealp{Giustini+2020:qpe2,Arcodia+2021:eroqpes,Arcodia+2024:qpe34}; hereafter \citetalias{Giustini+2020:qpe2,Arcodia+2021:eroqpes,Arcodia+2024:qpe34}), plus some candidates \citep{Chakraborty+2021:qpe5cand,Quintin+2023:tormund,Evans+2023:swift,Guolo+2023:swift}. QPEs were discovered for the most part (i.e., four out of six) by the eROSITA X-ray telescope (\citetalias{Arcodia+2021:eroqpes,Arcodia+2024:qpe34}). As the search method was purposely designed for the discovery of QPEs (\citetalias{Arcodia+2021:eroqpes,Arcodia+2024:qpe34}) and the eROSITA surveys are relatively homogeneous over the entire sky \citep{Merloni+2024:erass}, we are able to provide in this work, for the first time, an intrinsic volumetric rate of QPEs.
    
    Finally, most of the latest theoretical efforts interpret QPEs as triggered when the radiatively-efficient accretion disk of a massive black hole is repeatedly pierced by an orbiter of much smaller mass \citep[e.g.,][]{Xian+2021:qpemodel,Franchini+2023:qpemodel,Linial+2023:qpemodel2,Tagawa+2023:qpemodel,Zhou+2024:emri}\footnote{However, it is worth noting that alternative models exist (e.g. accretion disk instabilities, partial tidal disruptions; see \citetalias{Arcodia+2024:qpe34}).}. Intriguingly, this interpretation would make QPEs the electromagnetic counterparts of extreme mass ratio inspirals (EMRIs), which are detectable by future-generation GW detectors \citep[e.g.,][]{Amaro-Seoane+2018:emris}. Under the assumption that QPEs are indeed EMRIs, a remarkable implication is that the QPE volumetric rate would be the first-ever observationally-driven constraint to the EMRI rate. 
	
	\section{The search for QPEs with eROSITA}
	\label{sec:erosita}
	
	On 13 December 2019, eROSITA started observing in survey mode and completed four (and started the fifth) of the foreseen eight all-sky surveys (eRASS1 to eRASS8). In each survey, lasting six months, every source in the sky is observed for $\sim40\,$s (a `visit', hereafter) every $\sim4\,$h (i.e. a so-called eROday) for a total number of times within a single eRASS depending on its location in the sky. This number is typically around six on the ecliptic plane and increases towards higher ecliptic latitudes \citep{Merloni+2024:erass}. Most of the area in the sky has thus around $\approx6-10$ data points separated by $4\,$h, for a total baseline of $\sim1-2\,$d in the single eRASS \citep{Merloni+2024:erass} and about $\sim2-2.5\,$y for the eRASS1-eRASS4 (or eRASS1-eRASS5) baseline. Leading to the first QPE discoveries with eROSITA \citealp{Arcodia+2021:eroqpes,Arcodia+2024:qpe34}, our team has developed an algorithm to look for significant and repeated high-amplitude variability in the eROSITA sources. QPE candidates can be found within eROdays of the single eRASS or combining multiple eRASS. A first version of this method was briefly described in \citetalias{Arcodia+2021:eroqpes}. In brief, the algorithm searching for variable sources is primarily based on the X-ray light curves with no pre-screening on the counterpart nature. Then, more steps are taken in post-processing to classify each given source, such as the exclusion of secure Galactic objects using \emph{Gaia} DR3 \citep{GAIA+2016:gaia,GAIA+2021:edr3} and some further visual inspection of their X-ray and multi-wavelength properties for the handful of interesting cases per eRASS. Given eROSITA's cadence, X-ray follow-up with other X-ray telescopes is required for secure confirmation \citepalias{Arcodia+2021:eroqpes,Arcodia+2024:qpe34}. 
 
    Here, we present the QPE search method in more detail with the goal of providing the first estimate of the volumetric rate of QPEs. Light curves are extracted with the \texttt{srctool} task of the eROSITA Science Analysis Software System \citep[eSASS;][]{Brunner+2022:eSASS} from event files processing version 020, for all sources detected above detection likelihood \texttt{DET\_LIKE\_0} = 20 \citep{Merloni+2024:erass}. We exclude an area with a radius of three degrees around the South Ecliptic Pole, as it requires separate data processing and analysis \citep[e.g.,][]{Bogensberger+2024:sep}. Light curves are provided with 10-second bins between $0.2-0.6\,$keV, $0.6-2.3\,$keV and $2.3-5.0\,$keV. Light curves are thus rebinned to have a single count rate estimate per eROday \citepalias{Arcodia+2021:eroqpes,Arcodia+2024:qpe34}. This step is carried out by summing, within the eROday visit, the source and background counts in the unbinned light curve produced by eSASS. Then, we numerically compute the Poisson probability mass function (PPMF, with the \texttt{scipy} Python package \citealp{Virtanen+2020:SciPy}) for the count rate (CR) given the detected source and background counts, normalized by the areas of the extraction regions. We use the median value of the PPMF and $1\sigma$ percentiles as asymmetric upper (subscript $+$) and lower (subscript $-$) error bars. We compute the PPMFs for background alone and the total rate from source plus background. In the light curve extracted from the source area, visits compatible within uncertainties with background alone are considered undetected.
    
    An X-ray source is added to a pool of interesting alerts if it shows two (or more) visits which are significantly brighter than one (or more) fainter visit(s) in between. Our significance criteria are i) a ratio $>2$ between the lowest $1\sigma$ percentile of the PPMF of both bright eROday visits (dubbed HI; $CRHI - CRHI_{-}$) and the upper $1\sigma$ percentile of the fainter one in between (dubbed LO; $CRLO + CRLO_{+}$), and ii) maximum amplitude significance $>2$, computed as $AMP/ERR$ where $AMP = CRHI - CRHI_{-} - (CRLO + CRLO_{+})$ \citep[e.g.,][]{Boller+2022:amp} and $ERR=\sqrt{CRHI_{-}^2 + CRLO_{+}^2}$. On the order of $\approx550-650$ sources per eRASS satisfy these criteria, which are purposely loose as we aim to be complete, rather than pure, in finding QPEs from extreme variability from galaxies.
 
    For each eRASS, this preliminary sample is cross-matched with proper motion and parallax estimates to exclude Galactic objects. We match the X-ray coordinates with \emph{Gaia} DR3 \citep{GAIA+2016:gaia,GAIA+2021:edr3} using the python package \texttt{astroquery} \citep{Ginsburg+2019:astroquery} and store information of all \emph{Gaia} sources within 10". \emph{Gaia} proper motion (GPM) significance ($S_{GPM}$) is computed as $S_{GPM}$=GPM/$\sigma_{GPM}$, where GPM is computed from the RA and Dec proper motions as $\sqrt{GPM_{RA}^2 + GPM_{Dec}^2}$ and its uncertainty $\sigma_{GPM}=\sqrt{(\sigma_{GPM_{RA}}*GPM_{RA})^2 + (\sigma_{GPM_{Dec}}*GPM_{Dec})^2}/GPM$. Alerts associated with a single \emph{Gaia} source with proper motion significance above $5\sigma$ are excluded. The remaining alerts are typically $\approx 140-150$ per eRASS. Their X-ray and archival multi-wavelength properties are visually checked to exclude possible spurious Galactic contaminants which may be remaining after the first automated screening\footnote{Since a proper motion significance of $5\sigma$ is strict, several alerts can be easily identified as Galactic upon visualizing their optical images (e.g., The DESI Legacy Surveys\footnote{\href{https://www.legacysurvey.org/}{https://www.legacysurvey.org/}}; Aladin Sky Atlas, \citealp{Baumann+2022:aladin}) and archival photometric data \citep[e.g. via VizieR;][]{Ochsenbein+2000:vizier}.}, and others of extragalactic nature (e.g. known quasars and blazars). In addition, we also check alerts with no proper motion or parallax estimate in \emph{Gaia} ($\approx 70-100$ per eRASS). The vast majority of these consists of stars masked out from \emph{Gaia}, which can be easily excluded visualizing their optical images, but some local faint galaxies were also found to be missing a \emph{Gaia} estimate. After this pre-screening, the remaining alerts represent a subsample of likely-extragalactic X-ray sources. As mentioned above, our variability significance criteria are purposely loose, to favor detection completeness over purity. Hence, the algorithm triggers on stochastic variability of bright sources as well, which make the significance cut due to their smaller uncertainties. Apart from the extreme light curves like those of the eROSITA discovered QPEs \citepalias{Arcodia+2021:eroqpes,Arcodia+2024:qpe34}, the lower priority tiers most likely contains QPE sources with lower amplitude variability within eRASS, which cannot be easily told apart from quasar-like stochastic variability around the eRASS detection limit. We note that if a strikingly interesting light curve originates from a known AGN, this is not a exclusion criterion for QPE searches per se. However, typically these alerts are associated with fairly well-studied low-redshift sources, thus a closer look at their X-ray products from other (previous or following) eRASS and archival X-ray and multi-wavelength data is sufficient to exclude them as top-tier QPE candidates. A trivial case in which other eRASS can be used to assess the candidates is that of a given eRASS light curve showing seemingly high-amplitude variability (potentially indicating a QPE source), while other eRASS show a significant detection in most eROdays, but with insignificant variability. Cases in which the source is only detected in a single eRASS are instead kept in the pool of potential QPE candidates. Furthermore, a trivial case in which other X-ray or multi-wavelength archival data can be used to assess the candidates is that of well-studied local Seyfert galaxies with plenty of archival observations. We note that this step of visual inspection naturally imprints some degree of subjectivity in the choice of candidates. However, our best efforts were put into quantifying this as much as possible in our rates calculations (Sec.~\ref{sec:rates}) and supporting simulations (Appendix~\ref{sec:app}). After this screening, only a handful of highly-significant variable sources remain. Their archival multi-wavelength properties and photometry \citep[e.g. via VizieR;][]{Ochsenbein+2000:vizier} are explored for further analysis and follow-up. For instance, the variability criteria described here (namely, that significant high-amplitude variability repeated within the single eRASS) are the ones that found eRO-QPE1 and eRO-QPE2 \citepalias{Arcodia+2021:eroqpes}.
    
    In addition, we considered the case in which significant high-amplitude variability repeated across different eRASS, for instance with a single significant high state, but present in more than one eRASS. For this, the pool of alerts is obtained from sources that show one (or more) visit(s) significantly brighter than two (preceding and following) fainter ones. This single-flare search (in contrast with the repeated-flare version described above) yields $\approx 3500$ alerts per survey, of which $\approx 450$ remain after high significance proper motion is excluded (as flaring coronally emitting stars heavily dominate among the triggers, \citealp{Boller+2022:amp}), and another $\approx 450$ of sources with no proper motion estimate. Then, we take advantage of the presence of multiple eRASS and we select a subsample of sources which triggered the single-flare search in more than one eRASS (i.e. spreading the repeating variable behaviour across the different surveys rather than on the single one). This reduces the alerts to a handful of sources for further inspection. For instance, this is instead the method that found eRO-QPE3 \citepalias{Arcodia+2024:qpe34}. 
    
    Furthermore, after discovery of the first three QPEs we attempted a third set of criteria. Another subset of these $\approx 3500$ alerts, obtained with a single flare per eRASS, selected alerts in which all the faint visits are consistent with background and the signal is detected in one or two consecutive visits only, regardless of whether it significantly occurred in more than one eRASS or not. Effectively, this would consist in relaxing the repeating nature of the flare, but limiting to the highest significance of single flares per eRASS. This yields $\approx 200$ alerts per survey, of which $\approx 10$ remain after high significance proper motion is excluded, and another $\approx 30$ of sources with no proper motion estimate. The same visualization procedure mentioned above is applied to these alerts too. This is instead the method that found eRO-QPE4 \citepalias{Arcodia+2024:qpe34}.
    
    Finally, we note that all these searches have some obvious biases. For instance, they would likely miss potential QPE sources for which eruption last several days (i.e., longer than the baseline of the single eRASS) and repeat every tens of days, so that the source is alternatively detected or not in consecutive eRASS, but it is constant in each \citep[e.g.,][]{Evans+2023:swift,Guolo+2023:swift}. However, we note that the implementation of the single flare algorithm, crossmatched across the various eRASS, would allow eROSITA to detect short duration ($<12\,$h, namely the span of three eROdays) eruptions which repeat up to months. However, eROSITA has not found any such QPE sources (e.g., later confirmed to have recurrences over days-weeks and still durations over hours), yet. Perhaps, this suggests that the observed relation between average recurrence and average durations (e.g., \citep{Chakraborty+2021:qpe5cand}; Fig. 13 of \citetalias{Arcodia+2024:qpe34}) is intrinsic to the QPE population.

    \section{Intrinsic rates of QPEs}

    In this work, we aim to provide a first estimate of the intrinsic volumetric rate of QPEs. We adopt the four sources discovered by eROSITA \citepalias{Arcodia+2021:eroqpes,Arcodia+2024:qpe34} as representative of the intrinsic population and compute their detection efficiency (Section~\ref{sec:efficiency}) and resulting volumetric rates (Section~\ref{sec:rates}). We discuss our assumptions and their validity in Section~\ref{sec:assumpt} and Appendix~\ref{sec:app}.

    \begin{figure}[tb]
		\centering
		\includegraphics[width=0.99\columnwidth]{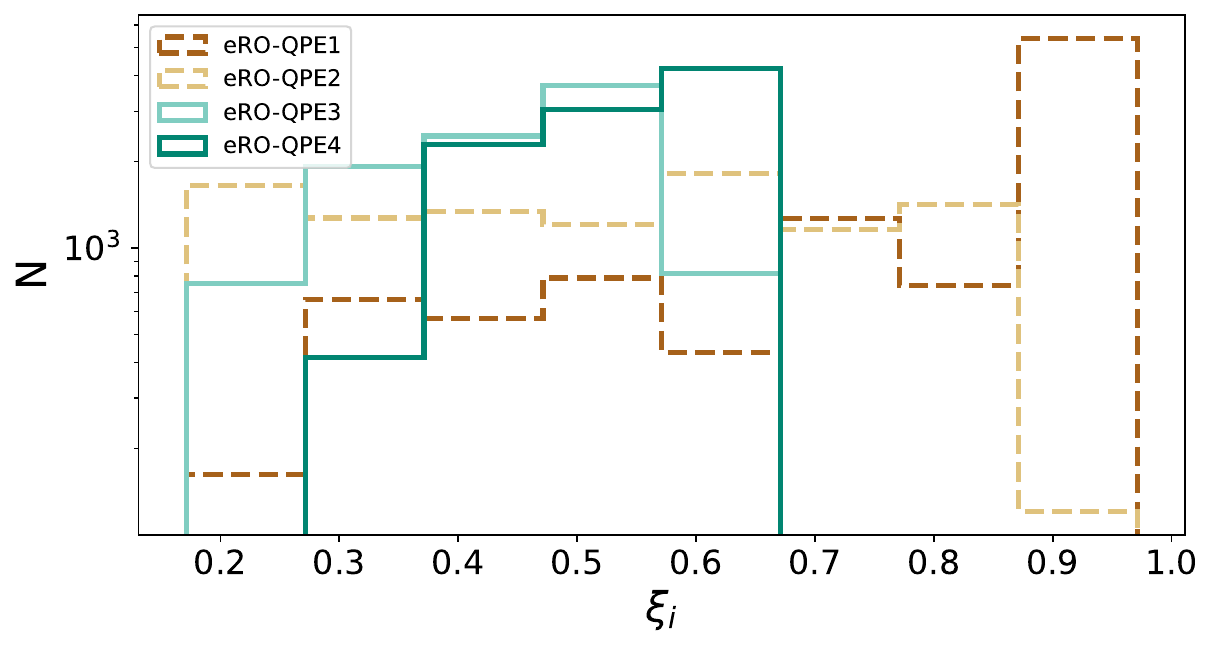}
		\caption{Histograms of the distribution of detection efficiency ($\xi$) of the QPE sources discovered by eROSITA. The sensitivity threshold required to significantly detect an eruption is drawn 10000 times from a Gaussian distribution. Each iteration of $\xi_i$ is then computed by downsampling the observed QPEs light curves through the real eRASS cadence and it represents the probability that our QPE search algorithm would trigger on the light curve of that given source.}
		\label{fig:hist}
	\end{figure}

    \subsection{Calculating the detection efficiency}
    \label{sec:efficiency}

    Estimating the detection efficiency of a given survey is a function of the peak luminosity distribution. Since QPEs may show appreciably different peak luminosity within the time span of a survey (\citetalias{Miniutti+2019:qpe1, Giustini+2020:qpe2,Arcodia+2021:eroqpes,Arcodia+2024:qpe34}; \citealp{Chakraborty+2021:qpe5cand}), we estimate their averaged values from available data. For eRO-QPE1, we use the average between the main peaks of the two \emph{XMM-Newton} observations \citepalias{Arcodia+2021:eroqpes}, which are representative of most of the time evolution shown by the source \citep{Chakraborty+2024:ero1}, namely $\langle L_{\rm 0.5-2.0\,keV}^{\rm peak}\rangle =1.2\times10^{43}\,$erg\,s$^{-1}$. For eRO-QPE2, the average peak luminosity in the \emph{XMM-Newton} observation presented in \citetalias{Arcodia+2021:eroqpes}: $\langle L_{\rm 0.5-2.0\,keV}^{\rm peak}\rangle =8.8\times10^{41}\,$erg\,s$^{-1}$. For eRO-QPE3, since the peak luminosity decayed over time \citepalias{Arcodia+2024:qpe34}, we adopted the average peak luminosity between the eRASS1 to eRASS4 epoch, $\langle L_{\rm 0.5-2.0\,keV}^{\rm peak}\rangle =9.7\times10^{41}\,$erg\,s$^{-1}$, when the source was actually detected by our algorithm. For eRO-QPE4, we used the average between the three \emph{XMM-Newton} peaks, namely $\langle L_{\rm 0.5-2.0\,keV}^{\rm peak}\rangle =4\times10^{42}\,$erg\,s$^{-1}$ \citepalias{Arcodia+2024:qpe34}.
   
    Regarding the detection efficiency, the key element is adopting the correct sensitivity of the searching algorithm. In searches for one-off transients that aim to simply detect a given source, the instrument sensitivity can be used. For our QPE search in particular (Section~\ref{sec:erosita}), we require the presence of a flux state somewhat brighter than the eRASS sensitivity in the single eROday and a specific pattern in the light curve, both in a single eRASS, or across eRASS (see \citetalias{Arcodia+2021:eroqpes,Arcodia+2024:qpe34}). Furthermore, visual inspection was necessary to confirm the presence of purely alternating variability patterns adding a subjective element which is not straightforward to quantify. Nonetheless, we identify eRO-QPE2 \citepalias{Arcodia+2021:eroqpes} as the QPE source with the lowest variability amplitude discovered so far with our search. In particular, we adopt as the QPE sensitivity threshold of a given visit the faintest count rate observed for eRO-QPE2 among the brightest that triggered the variability code (see Fig. 2 of \citetalias{Arcodia+2021:eroqpes}). This corresponds to a count rate of $0.43\pm0.12$ in the $0.6-2.3\,$keV band \citepalias{Arcodia+2021:eroqpes}. This count rate can be interpreted as the lowest eRASS count rate that a QPE bright state can have to be considered significantly variable according to our search method. Based on our experience, any variable source with a putative bright state fainter than this threshold, would not have been considered a reliable QPE candidate and would not have been followed up. Thus, we convert this eROSITA sensitivity observed flux to the corresponding count rate that would be observed by \emph{XMM-Newton} (for eRO-QPE2, eRO-QPE3 and eRO-QPE4) and \emph{NICER} (for eRO-QPE1), using \texttt{WebPIMMS}\footnote{\href{https://heasarc.gsfc.nasa.gov/cgi-bin/Tools/w3pimms/w3pimms.pl}{Link to WebPIMMS}}. We note that this count rate threshold, once converted to a flux with the spectral shape observed for each QPE source, also defines the maximum observability distance and the discovery volume of a given QPE source, given its observed average peak luminosity.
 
    For each of the discovered eROSITA QPEs, we adopt a mock eRASS sampling composed by as many eROdays as in two eRASS surveys of the actual light curves (see \citetalias{Arcodia+2021:eroqpes,Arcodia+2024:qpe34}). This filter, which mimics eROSITA's observational sampling during the survey, is then used to scan the \emph{XMM-Newton} and \emph{NICER} light curves of each confirmed QPE source. As an eROday visit exposure is $\sim40\,$s long, we shift this filter by $\sim40\,$s for enough times to cover at least one or a few full cycles of each QPE source. This time shift is adopted to be $\sim4.5\,$d for eRO-QPE1 (given the high scatter observed; \citetalias{Arcodia+2021:eroqpes}; \citealp{Chakraborty+2024:ero1}), $\sim5\,$h for eRO-QPE2, $\sim20.5\,$h for eRO-QPE3 and $\sim20\,$h for eRO-QPE4. If needed, the observed light curves are looped, ensuring that the observed average quasi-periodicity is maintained within the observed scatter. Then, we compute the detection efficiency as follows. For each scan with the eRASS mock filter, we consider it a QPE detection if at least two mock visits are found to be brighter than the count rate threshold (described above), with at least a visit in between which is below. Otherwise, the scan is considered unsuccessful. This directly compares with our actual QPE search (Sect.~\ref{sec:erosita}). The inferred efficiency, computed for each source separately, is therefore the ratio between the number of QPE detections and the total number of scans attempted with the filter. This efficiency can be interpreted as the probability that a given existing QPE source, with properties like those discovered, would trigger our search with a baseline of two eRASS. Since QPEs outlast the entire baseline given by all eRASS, we treat the two eRASS as consecutive for simplicity. We repeat the efficiency calculation 10000 times, drawing the count rate threshold from a Gaussian distribution given by the observed mean and standard deviation (Fig.~\ref{fig:hist}). We remind that this threshold is converted from the observed eRASS count rate of eRO-QPE2 with its $1\sigma$ uncertainty \citepalias{Arcodia+2021:eroqpes}. For a given QPE source ($QPE_n$), each draw \emph{i} provides a different efficiency ($\xi_i$, Fig.~\ref{fig:hist}) and a different maximum detectability luminosity distance ($d_{{\rm Lmax},i}$), volume ($V_{{\rm max},i}$, corrected by $\delta A=0.5$ since eROSITA-DE contains half of the entire sky) and redshift ($z_{{\rm max},i}$).

    \begin{figure}[tb]
		\centering
		\includegraphics[width=0.99\columnwidth]{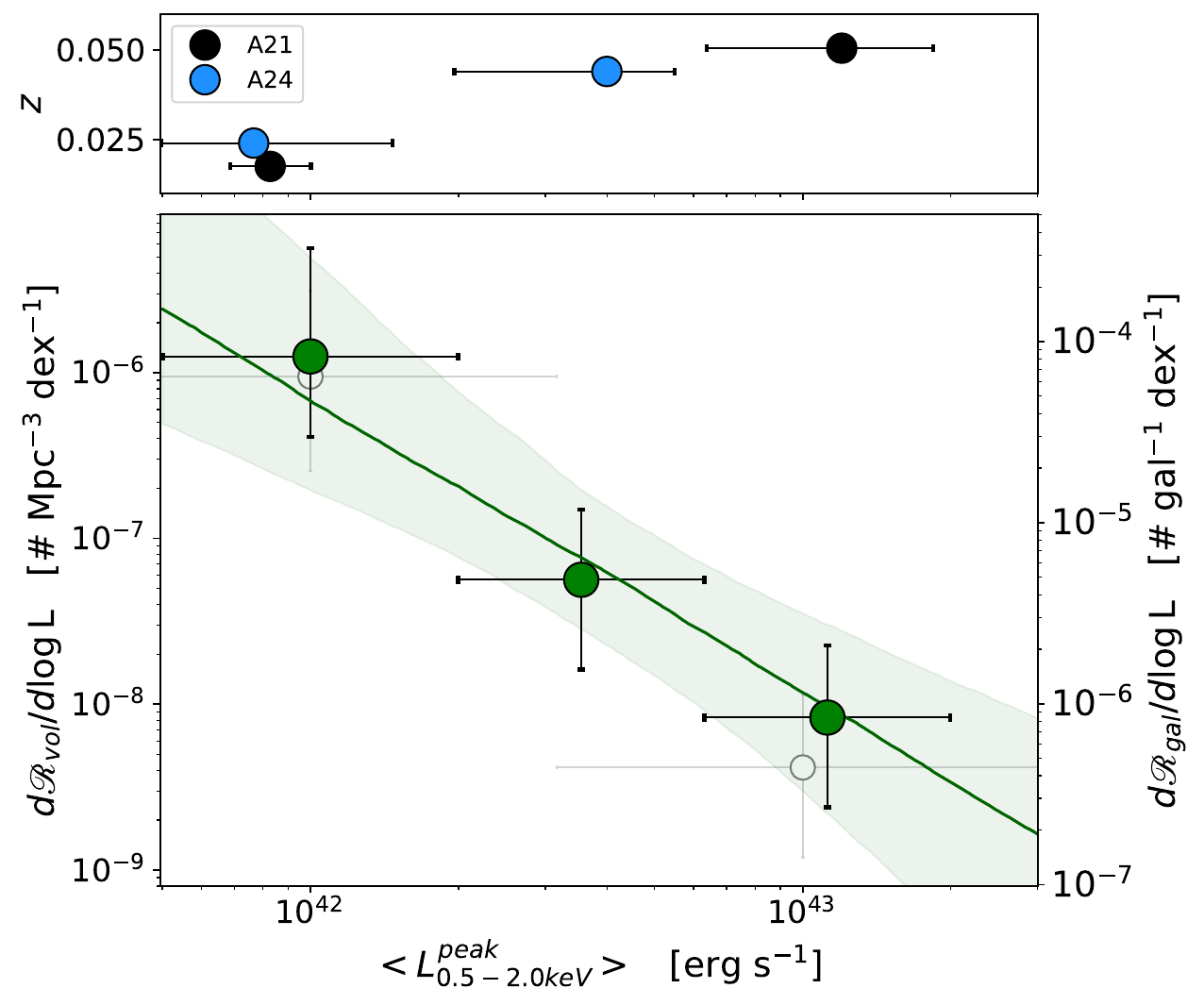}
		\caption{\emph{Top}: the distribution of the four eROSITA QPEs (\citetalias{Arcodia+2021:eroqpes,Arcodia+2024:qpe34}) in the luminosity-redshift plane. \emph{Bottom}: X-ray luminosity function of eROSITA QPEs (green), expressed both in Mpc$^{-3}$ and per galaxy, as a function of the average peak $0.5-2.0\,$keV luminosity. The three bins adopted in $\langle \log L_{\rm 0.5-2.0\,keV}^{\rm peak}\rangle$ are $41.7-42.3$, $42.3 - 42.8$ and $42.8 - 43.3$, containing two, one and one QPEs, respectively. The green line and related contours represent the median and 16-84th percentiles of the fit power-law model, with a slope $\approx - 2$. The two grey points represent the rates from eRO-QPE1 and eRO-QPE2 alone (\citetalias{Arcodia+2021:eroqpes}) in two bins of 1\,dex, to highlight the small effect of adding new discoveries.}
		\label{fig:rates}
	\end{figure}
    
    \subsection{Calculating volumetric rates}
    \label{sec:rates}
    
    We adopt three bins in $\langle \log L_{\rm 0.5-2.0\,keV}^{\rm peak}\rangle$, namely $41.7-42.3$, $42.3 - 42.8$ and $42.8 - 43.3$. The lowest bound of 41.7 is chosen as it is the lowest peak flux shown in the eRASS light curves by any eROSITA-discovered QPE source. This choice is also supported by the simulations reported in Appendix~\ref{sec:app}, since the eRASS detection efficiency is zero below this luminosity even in the low-z Universe. These bins contain two, one, and one QPE sources, respectively. For each luminosity bin $\Delta \log L_j$ we compute the volumetric rate $\mathscr{R}_{{\rm vol},j}$ by maximizing the mean Poisson likelihood obtained from the distribution of 10000 iteration, each giving, given the rate and $k$ detections, $p_i(k|\mathscr{R}_{{\rm vol},j}) = e^{-\lambda_i} \lambda_i^k / k!$, where $\lambda_i=\mathscr{R}_{{\rm vol},j}\,\xi_i\,V_{{\rm max},i}\, \delta A$. This calculation approximates the distribution of QPE sources over luminosity and redshift with the known detected eROSITA QPEs, the total volume with $V_{{\rm max}}$ and that the nuisance parameters (e.g., $\xi_i$ and $V_{{\rm max},i}$) are distributed according to the count rate measurement uncertainties used as sensitivity. The $1\sigma$ uncertainties on the mean Poisson distribution are taken at the points where the likelihood drops by a factor of $e^{-0.5}$. 
    We plot the resulting X-ray luminosity function of QPEs in the bottom panel of Fig.~\ref{fig:rates}. The top subpanel of Fig.~\ref{fig:rates} shows instead the redshift-luminosity plane, where the luminosity uncertainties reflect the diversity in peak luminosity observed rather than the statistical uncertainty on an individual peak estimate. Comparing with the luminosity function of TDEs \citep{Yao+2023:TDEs} we note a similar decrease of a factor $\approx100$ per luminosity decade. We fit a power-law relation through the luminosity function of Fig.~\ref{fig:rates}, with uncertainties on both axes, using \texttt{Ultranest} \citep{Buchner2021:ultranest}, with $\log \Phi = \alpha_0 + \beta_0 * (\log L^{\rm peak}_{\rm 0.5-2.0\rm\,keV} - 42)$, where $\Phi = d \mathscr{R}_{{\rm vol}} / d \log L$. We obtained a slope with median (and 16th-84th percentiles) of $\beta_0 = -1.83^{+0.73}_{-1.03}$, with a normalization $\alpha_0 = -6.16^{+0.85}_{-0.54}$. We show the median (and 16th-84th percentile contours) power-law function in green in Fig.~\ref{fig:rates}. Integrating this function within its uncertainties above $\log L_{0.5-2.0\,keV}^{\rm peak} > 41.7$ we obtain a median (and related 16th-84th percentiles) volumetric abundance rate of $\mathscr{R}_{{\rm vol}} = 0.60_{-0.43}^{+4.73} \times 10^{-6}\,$Mpc$^{-3}$.
        
    We assume a volumetric density of galaxies with a stellar mass within $10^{8.5-10.5}$, which includes that of the known eROSITA QPEs (\citetalias{Arcodia+2021:eroqpes,Arcodia+2024:qpe34}), of $\sim 1.65 \times 10^{-2}\,{\rm Mpc}^{-3}$ \citep{Baldry+2012:GAMA}. Therefore, our volumetric abundance rate corresponds to a per-galaxy abundance rate of $\mathscr{R}_{\rm gal} = 0.36_{-0.26}^{2.87} \times 10^{-4}\,$gal$^{-1}$, above an intrinsic average $\log L_{0.5-2.0\,keV}^{\rm peak} > 41.7$. This implies that at any time, roughly one in 10000 galaxies in that stellar mass range is erupting quasi-periodically. Relating this to a formation rate relies on assumptions on the typical QPE lifetime. Each secure QPE source as been emitting eruptions since discovery \citep[even if perhaps not continuously,][]{Miniutti+2023:gsnreapp}, up to more than 20 years \citepalias{Giustini+2020:qpe2}. Therefore, we assume a typical lifetime of $\tau_{\rm life}\approx10$ years as an example, but leave it as a free parameter. This yields a formation rate $\mathscr{R}_{\rm gal}/\tau_{\rm life}$ of approximately a $\approx 0.4 \times 10^{-5}(\tau_{\rm life}/10\,\mathrm{y})^{-1}$ per galaxy per year or $\mathscr{R}_{\rm vol}/\tau_{\rm life}\approx 0.6 \times 10^{-7}(\tau_{\rm life}/10\,\mathrm{y})^{-1}$ per Mpc$^3$ per year.

    \subsection{Assumptions and approximations}
    \label{sec:assumpt}
    
    Since the overall QPE activity, excluding potential short phases with disappearance \citep{Miniutti+2023:gsnreapp}, appears to last years to decades (\citetalias{Miniutti+2019:qpe1, Giustini+2020:qpe2,Arcodia+2021:eroqpes,Arcodia+2024:qpe34}; \citealp{Miniutti+2023:gsnreapp}), QPEs outlast a given survey. Therefore, observed rates can be directly compared to abundance volumetric rates, whilst inferring formation rates strongly relies on the unknown typical lifetime of QPEs. The calculation of intrinsic abundance rates itself is also affected by two major factors. One is astrophysical, namely the knowledge of the intrinsic QPE population, its intrinsic peak luminosity, duration and recurrence distributions and if and how they relate. The other is observational, namely the detection efficiency of a given instrument. Estimating the latter is possible with eROSITA, given its relatively homogeneous survey strategy \citep{Merloni+2024:erass} and our detection algorithm purposely designed to search for QPE-like flares (see Section~\ref{sec:efficiency}). The former is by nature unknown and it is particularly difficult for QPEs given the wide diversity in timing properties (i.e. duration and recurrence of eruptions and their evolution over time) within only a handful of known sources (\citetalias{Miniutti+2019:qpe1, Giustini+2020:qpe2,Arcodia+2021:eroqpes,Arcodia+2024:qpe34}; \citealp{Chakraborty+2021:qpe5cand,Arcodia+2022:ero1_timing,Miniutti+2023:gsnreapp,Miniutti+2023:gsnrebr,Chakraborty+2024:ero1}; Giustini et al., in prep.). 
    
    In this work, we adopt the properties of the four known eROSITA QPEs as representative of the population. The arguments presented in the previous sections had the objective to provide a (qualitative) validation of this approach. For instance, eRO-QPE2 \citepalias{Arcodia+2021:eroqpes} and eRO-QPE3 \citepalias{Arcodia+2024:qpe34} have compatible average peak X-ray luminosity, but widely different burst durations ($\sim0.5\,$h and $\sim3$\,h, respectively) and recurrences ($\sim2.4\,$h and $\sim20$\,h, respectively). Most importantly, they reside in the most intrinsically populated luminosity bin. Therefore, as long as the burst properties of eRO-QPE2 and eRO-QPE3 span the bulk of the intrinsic population, we do not expect to dramatically underestimate or overestimate the intrinsic rates provided here. Furthermore, none of the QPE sources discovered so far seem to have a severely skewed efficiency distribution towards high values (Fig.~\ref{fig:hist}), thus eROSITA did not select only the sources which are the easiest to find with its sampling. More quantitatively, we verified these assumptions a posteriori, by simulating QPE light curves with various peak luminosity, recurrence and duration of the eruptions. We computed the eRASS detection efficiency of these mock light curves and concluded that, based on the current knowledge on QPEs, the known eROSITA-discovered QPE sources are not a significantly biased draw from the intrinsic population. Hence, they can indeed be used for a meaningful first estimate of their intrinsic rate. We report more details on the simulations in Appendix.~\ref{sec:app}.
    
    Nevertheless, several complications remain, that may affect the calculations presented in this work. In particular, the discovery rate of QPEs in the eROSITA data is, to some extent, relying on subjective criteria based on visual inspection, as we discussed at length in Section~\ref{sec:erosita}. However, we note that eRO-QPE1 and eRO-QPE2 were the only two QPEs found with a single consistent method \citepalias{Arcodia+2021:eroqpes} using only eRASS1 and eRASS2 data. However, performing the same calculation above for eRO-QPE1 and eRO-QPE2 with two luminosity bins of one dex (41.5-42.5 and 42.5-43.5) seems to yield consistent results (grey points in Fig.~\ref{fig:rates}). Finally, we note that the inferred rates are not corrected for X-ray absorption. Known QPE sources are rather unabsorbed (\citetalias{Miniutti+2019:qpe1, Giustini+2020:qpe2,Arcodia+2021:eroqpes,Arcodia+2024:qpe34}; \citealp{Chakraborty+2021:qpe5cand}), with only eRO-QPE2 being absorbed by a moderate column density ($N_H\sim3\times 10^{21}\,$cm$^{-2}$; \citetalias{Arcodia+2021:eroqpes}). As a rough estimate, one may consider that the current rates underestimate the intrinsic QPE population by a factor $\lessapprox2$, following results of the fraction of obscured nuclear super-massive black holes \citep[e.g.,][]{Buchner+2015:obscur,Carroll+2023:cthick}.

\section{Summary and future prospects}

X-ray quasi-periodic eruptions are the most recent addition to the group of extragalactic nuclear transients. In this work, we report the first volumetric abundance rate of QPEs, computed with the first four eROSITA discoveries (\citetalias{Arcodia+2021:eroqpes,Arcodia+2024:qpe34}). We obtained $\mathscr{R}_{{\rm vol}} = 0.60_{-0.43}^{+4.73} \times 10^{-6}\,$Mpc$^{-3}$ above an intrinsic average $\log L_{0.5-2.0\,keV}^{\rm peak} > 41.7$, or $\mathscr{R}_{\rm gal} = 0.36_{-0.26}^{2.87} \times 10^{-4}\,$gal$^{-1}$. This yields a formation rate $\mathscr{R}_{\rm gal}/\tau_{\rm life}\approx 0.4 \times 10^{-5}(\tau_{\rm life}/10\,\mathrm{y})^{-1}$\,gal$^{-1}$\,year$^{-1}$ or $\mathscr{R}_{\rm vol}/\tau_{\rm life}\approx 0.6 \time 10^{-7} (\tau_{\rm life}/10\,\mathrm{y})^{-1}$\,Mpc$^{-3}$\,year$^{-1}$, which, however, depends on the unknown QPE lifetime $\tau_{\rm life}$.

\begin{figure}[tb]
		\centering
		\includegraphics[width=0.87\columnwidth]{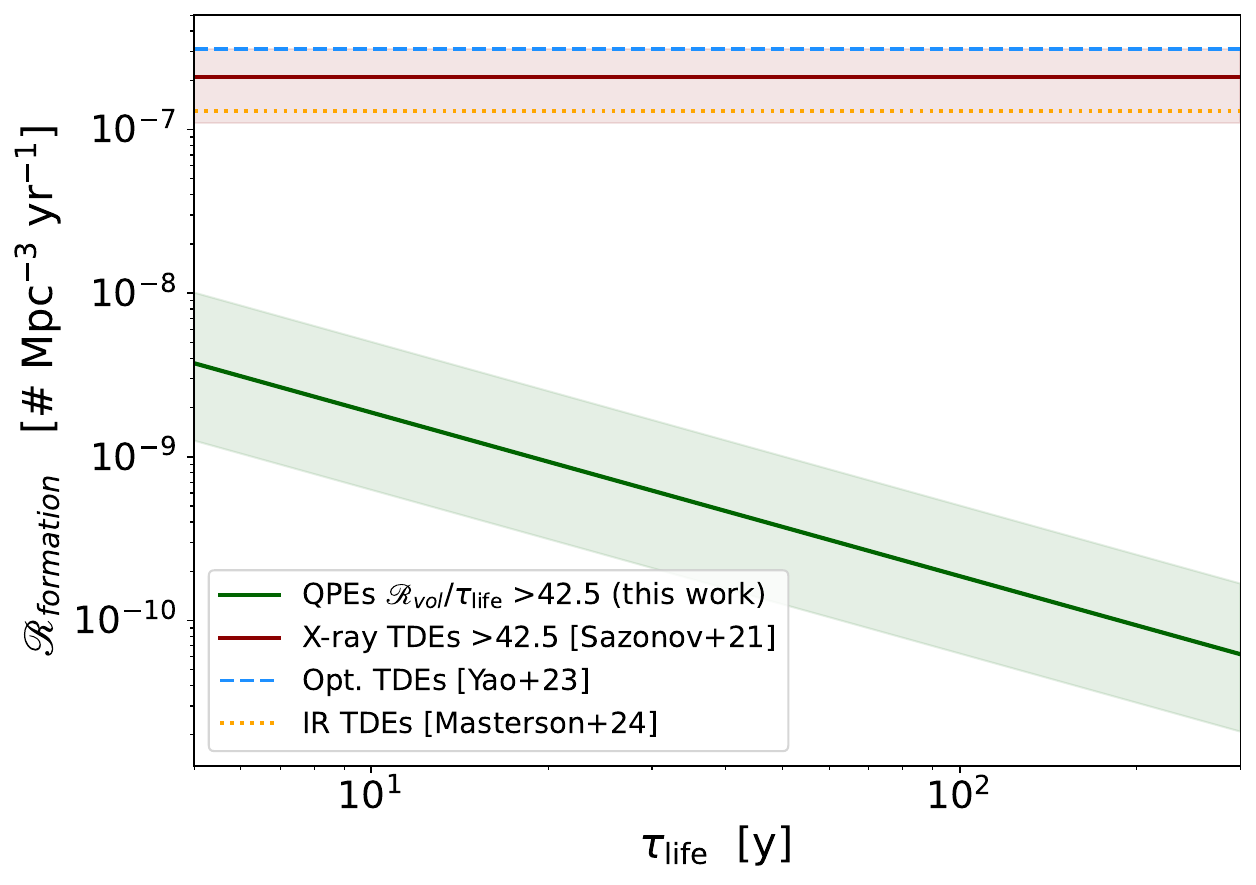}
		\caption{The QPE formation rate $\mathscr{R}_{\rm formation}$ as a function of $\tau_{\rm life}$ is shown with a green line and contour, for median and $1\sigma$ range, respectively. It is computed above $\log L_{0.5-2.0\,keV}^{\rm peak} > 42.5$ for direct comparison with the X-ray TDE formation rate from \citet{Sazonov+2021:erotde}, shown in dark red. TDE rates from optical \citep{Yao+2023:TDEs} and infrared \citep{Masterson+2024:irtde} searches are also shown, as shown in the legend. Within the ``QPE = TDE + EMRI'' scenario \citep[e.g.,][]{Franchini+2023:qpemodel,Linial+2023:qpemodel2}, on the order of $\approx 0.01\,(\tau_{\rm life}/10\,\mathrm{y})^{-1}$ of galaxies with TDEs could later on show signatures of QPEs.}
		\label{fig:rates_comparison_TDE}
\end{figure}

This formation rates agrees to zeroth-order with that inferred by some of the proposed models on QPEs triggered by secondary orbiters \citep[e.g.,][]{Zhao+2022:ptde,Lu&Quataert2023:ptde,Linial+2023:qpemodel2}, naturally within the large uncertainties and variables in play. However, the qualitative predictions on the emission mechanism and resulting luminosity of these modeled systems prevent us from making more quantitative comparisons with our observed rates, computed above $\log L_{0.5-2.0\,keV}^{\rm peak} > 41.7$. Here, we instead attempt a more agnostic comparison between QPE rates and those of other extragalactic nuclear transients like TDEs. Given the apparent dichotomy \citep[e.g.,][]{Malyali+2023:dichotomy,Guolo+2023:tdes} between TDEs discovered in the optical \citep[e.g.,][]{Yao+2023:TDEs} and X-rays \citep[e.g.,][]{Sazonov+2021:erotde}, now with the further latest discoveries via infrared selection \citep{Masterson+2024:irtde}, this comparison requires particular care. The most direct, thus reliable, comparison can be performed with the X-ray TDE rate from \citet{Sazonov+2021:erotde}. Thus, we integrate our QPEs luminosity function (Fig.~\ref{fig:rates}) starting from the same value ($\log L_{0.5-2.0\,keV}^{\rm peak} > 42.5$) and obtain a formation rate $\mathscr{R}_{\rm formation}=\mathscr{R}_{\rm vol}/\tau_{\rm life}\approx 2 \times 10^{-9} (\tau_{\rm life}/10\,\mathrm{y})^{-1}$\,Mpc$^{-3}$\,year$^{-1}$. We plot $\mathscr{R}_{\rm formation}$ as a function of $\tau_{\rm life}$ in Fig.~\ref{fig:rates_comparison_TDE} together with the X-ray TDEs value from \citet{Sazonov+2021:erotde}. Based on Fig.~\ref{fig:rates_comparison_TDE}, and assuming the ``QPE = TDE + EMRI'' scenario \citep[e.g.,][]{Franchini+2023:qpemodel,Linial+2023:qpemodel2},
this would imply that on the order of $\approx 0.01\,(\tau_{\rm life}/10\,\mathrm{y})^{-1}$ of galaxies with (X-ray bright) TDEs could later on show signatures of QPEs. Integrated formation rates of optically- \citep{Yao+2023:TDEs} and infrared-selected TDEs \citep{Masterson+2024:irtde} agree with the X-ray TDEs estimate, although they should be compared with caution given the complex time-evolving and inhomogeneous multi-wavelength properties of TDEs \citep{Guolo+2023:tdes,Masterson+2024:irtde}. This prediction can be tested as future larger sample of confirmed X-ray TDEs and QPEs are unveiled, provided systematic and homogeneous follow-up campaigns are performed at late times.

\subsection{Predictions for LISA}

\begin{figure}[tb]
		\centering
		\includegraphics[width=0.87\columnwidth]{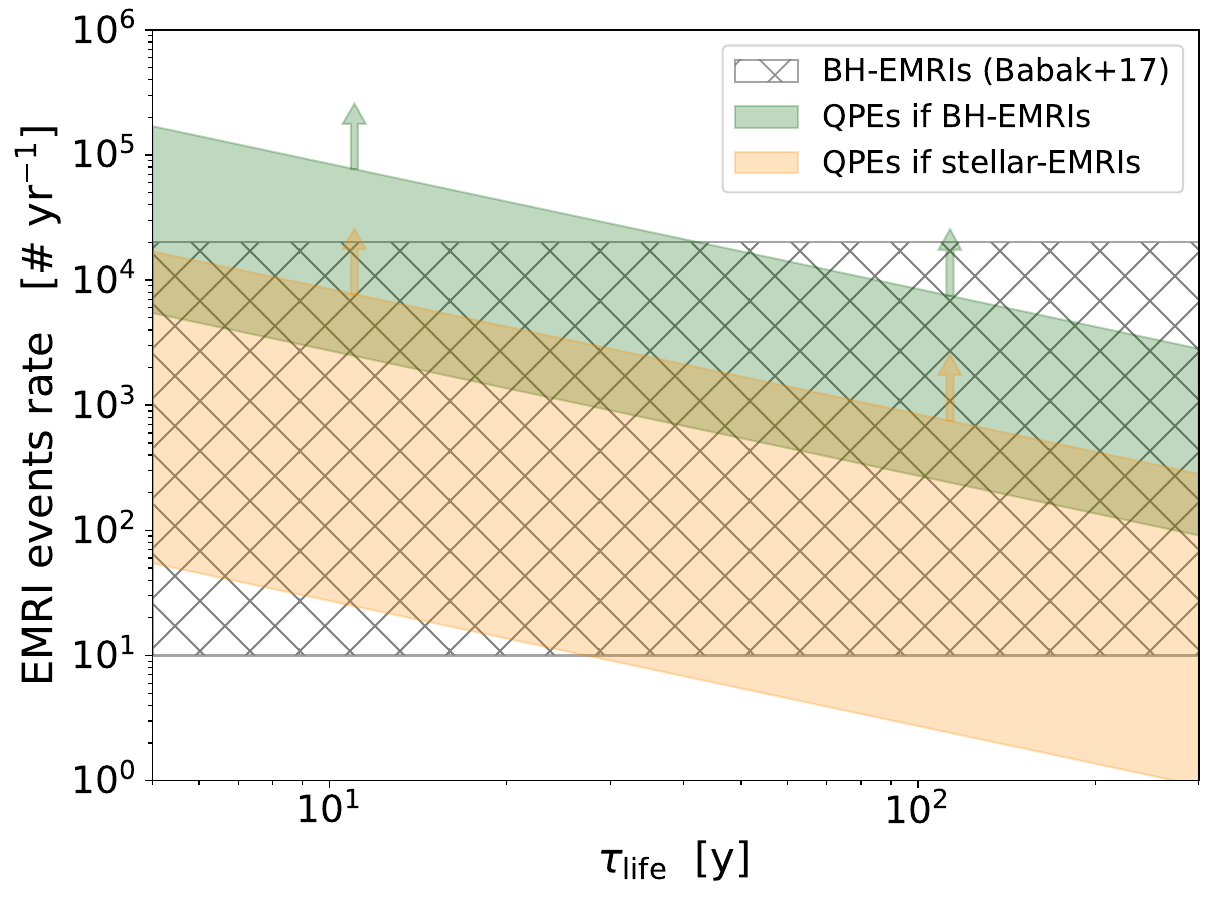}
		\caption{Event rates of BH-EMRIs within $z=1$. Predicted theoretical EMRI rates (hatched grey area) from \citet{Babak+2017:emris} are compared to those inferred from the QPE formation rate, which have to be interpreted as lower limits as not all EMRIs, even if QPEs are such, are expected to emit eruptions. In case QPEs are BH-EMRI, a direct comparison can be made (green $1\sigma$ interquantile range), while if QPEs are stellar-EMRIs the constraint for BH-EMRIs (orange $1\sigma$ interquantile range) requires further extrapolations. Nonetheless, and particularly in the former case, pessimistic EMRI population models would be disfavored, with huge implications for LISA detections.}
		\label{fig:rates_comparison_EMRI}
\end{figure}

In case the EMRI scenario will be unambiguously confirmed as the origin for QPEs, the volumetric rate estimate presented in this work would be the first-ever observational constraint on the EMRI rate, which is of paramount importance for future-generation GW detectors like the Laser Interferometer Space Antenna (LISA) and Tianqin \citep{Amaro-Seoane+2018:emris}. Since not all EMRIs are expected to emit QPEs, the QPE rate would naturally be a lower limit. Even if statistically every EMRI is likely to experience an independent TDE that throws gas on it with the potential of revealing it \citep{Linial+2023:qpemodel2}, the unknown modulo is the fraction of EMRIs for which the secondary is of the nature (a star, and of which kind, or a black hole), mass and orbital configuration required to produce observable X-ray eruptions. Currently, even within the proposed EMRI models it is still debated whether the orbiter is a star \citep{Xian+2021:qpemodel,Linial+2023:qpemodel2,Tagawa+2023:qpemodel}, a black hole \citep{Franchini+2023:qpemodel}, or even whether it can be both. Hence, we only attempt to provide some useful values under the assumption that QPEs are EMRIs, and we start assuming they are black hole EMRIs for simplicity. We start by providing the abundance rate of QPEs per massive black hole (MBH, $\mathscr{R}_{\rm MBH}$) integrating the mass function from \citet{Merloni+2008:bhmf} in the range $\log M_{BH} = 5.0-7.5$, which yields $\mathscr{R}_{\rm MBH} = 0.32_{0.23}^{2.56} \times 10^{-4}\,$MBH$^{-1}$. Hence, there are at least as many EMRIs per MBH, of which LISA is expected to detect a fraction of $\sim(10-40)\%$ \citep{Babak+2017:emris}. In particular, the EMRI population models predict a cosmic rate of EMRIs in units of $\sim(10-2 \times 10^4)\,$year$^{-1}$ \citep{Babak+2017:emris}, thus we integrate the QPE abundance rate $\mathscr{R}_{{\rm vol}}$ up to $z=1$ \citep[where most EMRI population models are complete in terms of LISA detection rates;][]{Babak+2017:emris}, using the cosmology from \citep{Hinshaw+2013:wmap9}, and obtain a $1\sigma$ interquantile range of $(3-85)\times 10^3\,(\tau_{\rm life}/10\,\mathrm{y})^{-1}$\,year$^{-1}$. We show this lower limit prediction in green in Fig.~\ref{fig:rates_comparison_EMRI}. Conversely, in case QPEs are stellar-EMRIs, we can attempt to scale their rate to that of black hole EMRIs. Using the framework of \citet{Linial+2022:segreg,Linial+2023:unstable} and assuming a $10\%$ binary fraction at the influence radius and that the relative abundance of stellar-mass black holes and stars in the nuclear star cluster is $10^{-3}$, the rate of the former is $\approx (1-10)\%$ of the latter (I. Linial, priv. comm.). This implies that the $1\sigma$ interquantile range of the lower limit to the rate of EMRIs detectable by LISA is in the range $\approx (30-8500)\,(\tau_{\rm life}/10\,\mathrm{y})^{-1}$\,year$^{-1}$. We show this lower limit prediction in orange in Fig.~\ref{fig:rates_comparison_EMRI}. Both estimates of stellar and black hole EMRIs disfavor the pessimistic population models \citep{Babak+2017:emris}, shown as a grey hatched area in Fig.~\ref{fig:rates_comparison_EMRI}. In particular, the QPE rate lower limit would be quite stringent if QPEs are the electromagnetic counterparts of black hole EMRIs.

\subsection{Predictions for future-generation X-ray observatories}
 
As a by-product of the inferred volumetric rates, we can provide predictions for future-generation X-ray telescopes for the first time. So far, known QPEs have flared in the soft X-ray band only\footnote{The disk-collision framework predicts the possible existence of UV QPEs \citep{Linial+2023:uv}. If this were to be the case, future UV missions like ULTRASAT \citep{Sagiv+2014:ultrasat} or UVEX \citep{Kulkarni+2021:uvex} may also discover QPE sources.} (with most signal below 2\,keV) and, as we have found in this work, their abundance rates are relatively low. Therefore, a soft X-ray telescope with a large effective area and a large field of view is desirable to systematically discover QPEs. In practice, as we have shown with our eROSITA search (Sect.~\ref{sec:erosita}), in order to significantly resolve an eruption and recognize it as such within the data stream of a given telescope, the peak flux has to be somewhat brighter than the instrument's sensitivity. For simplicity, we adopt this eruption-detection threshold as three times the nominal sensitivity of a given instrument and assume a volumetric rate of $\sim0.6 \times 10^{-6}\,$Mpc$^{-3}$ at $L_{x,{\rm peak}}\sim 10^{42}\,$erg\,s$^{-1}$. 

For instance, the wide-field X-ray telescope (WXT) on the upcoming Einstein Probe (EP) mission \citep{Yuan+2022:EP} boasts a large instantaneous field of view, although its sensitivity is only tens of times shallower than the brightest QPE peak observed so far. Therefore, simulations of stacked images are required to make further predictions. Instead, we find that the Wide Field Imager \citep{Meidinger+2020:wfi} onboard Athena \citep{Nandra+2013:athena} may detect $\approx2$ QPE sources per 15\,deg$^2$ survey. Whether these sources are immediately discovered as QPE emitters or not depends on the exposure of the survey ($t_{\rm exp}$) compared to the QPE duty cycle or recurrence time ($t_{\rm recur}$), such as a factor $\eta = t_{\rm exp}/t_{\rm recur}$. With a typical recurrence time of $\approx10$\,h\,$=36\,$ks, a single moderately deep 10\,ks  15\,deg$^2$ or a wider (100\,deg$^2$) and shallower (1.5\,ks) survey would both discover of order unity QPE sources. Here, we adopted a WFI sensitivity of (three times) $\sim10^{-15}\,$erg\,s$^{-1}$\,cm$^{-2}$ in $1\,$ks from \citet{Piro+2022:athena}. 

Finally, we note that for an efficient QPE discovery machine, repeating the same survey in the same area for a few times is desirable, as it would compensate the factor $\eta$ after a number $N_s\sim 1/\eta$ of surveys. However, performing more than $N_s$ would not be effective as eruptions from the same QPE source would be detected, rather than new QPE sources. Thus, an ideal QPE discovery machine would move to a different sky area after $N_s\sim 1/\eta$ surveys and complete as much of the entire sky as possible. 	


	%
	%
\bibliographystyle{aa} 
\bibliography{bibliography} 

\begin{appendix}        
\section{Simulating light curves of an empiric population of QPE sources}
\label{sec:app}

\begin{figure}[tb]
		\centering
		\includegraphics[width=0.93\columnwidth]{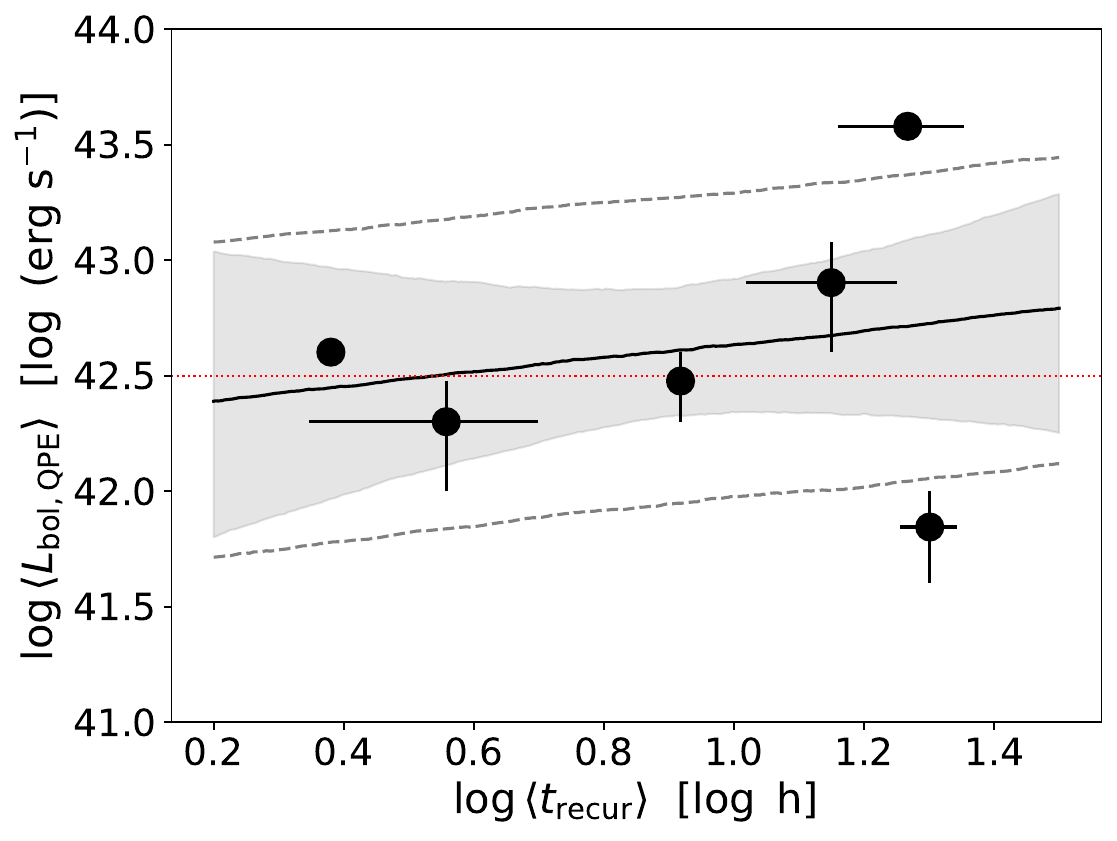}
        \includegraphics[width=0.95\columnwidth]{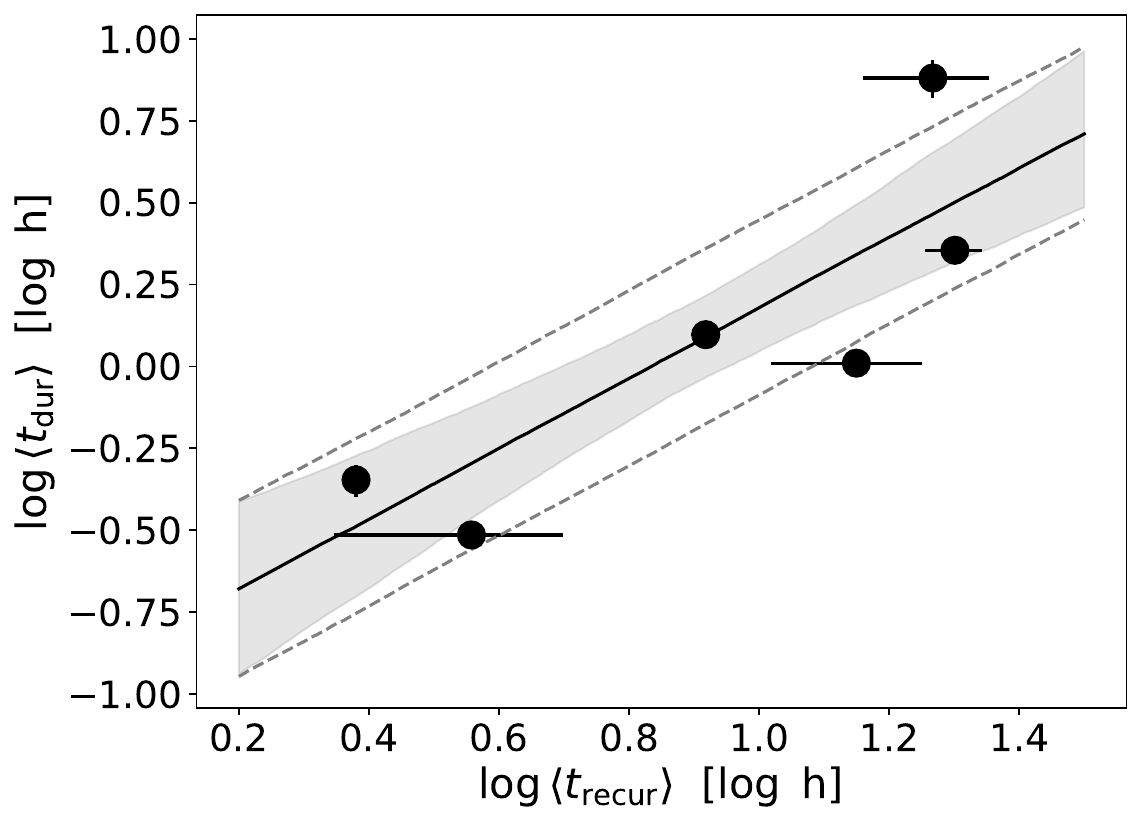}
		\caption{Relation between average recurrence time ($t_{\rm recur}$) and average peak X-ray luminosity (\emph{top panel}) or average eruption duration ($t_{\rm dur}$, \emph{bottom panel}) in QPE sources. Data from Fig.~13 of \citetalias{Arcodia+2024:qpe34}. In the top panel, the slope is consistent with zero (red dotted line), thus with quantities being uncorrelated given available data. In the bottom panel, the slope is consistent with being linear in the log-log plane.}
		\label{fig:time_fit}
\end{figure}

As discussed in Section~\ref{sec:assumpt}, the accuracy of the intrinsic rates provided in this work relies on the assumption that the known QPE sources discovered by eROSITA so far \citepalias{Arcodia+2021:eroqpes,Arcodia+2024:qpe34} are fairly representative of the intrinsic observable QPE population. In particular, on the assumption that the eROSITA detection efficiency probed by the known QPEs is not skewed towards larger values compared to that of the intrinsic population. Here, we verified that this assumption is indeed reasonable, by simulating QPE light curves and measuring their detection efficiency. The efficiency of any survey depends on the peak luminosity and that of eROSITA, given its sampling \citep{Merloni+2024:erass}, depends on the duration and recurrence of eruptions as well. In this work, we choose to simulate an empirical QPE population based on the data currently available. In particular, fitting data from \citetalias{Arcodia+2024:qpe34} we obtained a lack of correlation between average peak luminosity and average recurrence time. The fitted log-linear relation (top panel of Fig.~\ref{fig:time_fit}) shows a slope consistent with zero (namely $0.3\pm0.7$). Hence, at least for luminosity and recurrence we adopt an agnostic 2D grid with no correlation between the two quantities. However, we take advantage from the fact that there seem to be a relation between the average recurrence and duration across QPE sources (see the bottom panel of Fig.~\ref{fig:time_fit} and \citetalias{Arcodia+2024:qpe34}). We fit a linear relation between the logarithm of average recurrence and duration, using the values reported in \citetalias{Arcodia+2024:qpe34}. We obtain a normalization of $-0.90^{+0.34}_{-0.38}$, a slope of $1.08^{+0.36}_{-0.35}$, and intrinsic scatter $0.29^{+0.16}_{-0.10}$ (we show this relation in the bottom panel of Fig.~\ref{fig:time_fit}). Hence, we generated the QPE population drawing from a grid in peak luminosity ($\langle \log L_{\rm 0.5-2.0\,keV}^{\rm peak}\rangle$) and average recurrence ($\langle t_{\rm recur}\rangle$), the former ranging between $41-44\, \log($erg\,s$^{-1})$ and the latter between $1-30\,$h. These values bracket those of known QPE sources. A scatter of $10\%$ is assumed on both quantities, which is a representative value based on all the secure QPE sources \citepalias{Miniutti+2019:qpe1, Giustini+2020:qpe2,Arcodia+2021:eroqpes,Arcodia+2024:qpe34}. This scatter is not only observed in those discovered by eROSITA, thus it is not a bias due to eROSITA's cadence. Then, for each value of $\langle t_{\rm recur}\rangle$, we extract a value for the average QPE duration ($\langle t_{\rm dur}\rangle$) from the relation in the bottom panel of Fig.~\ref{fig:time_fit}, within its $1\sigma$ uncertainties. 

\begin{figure}[tb]
		\centering
		\includegraphics[width=0.99\columnwidth]{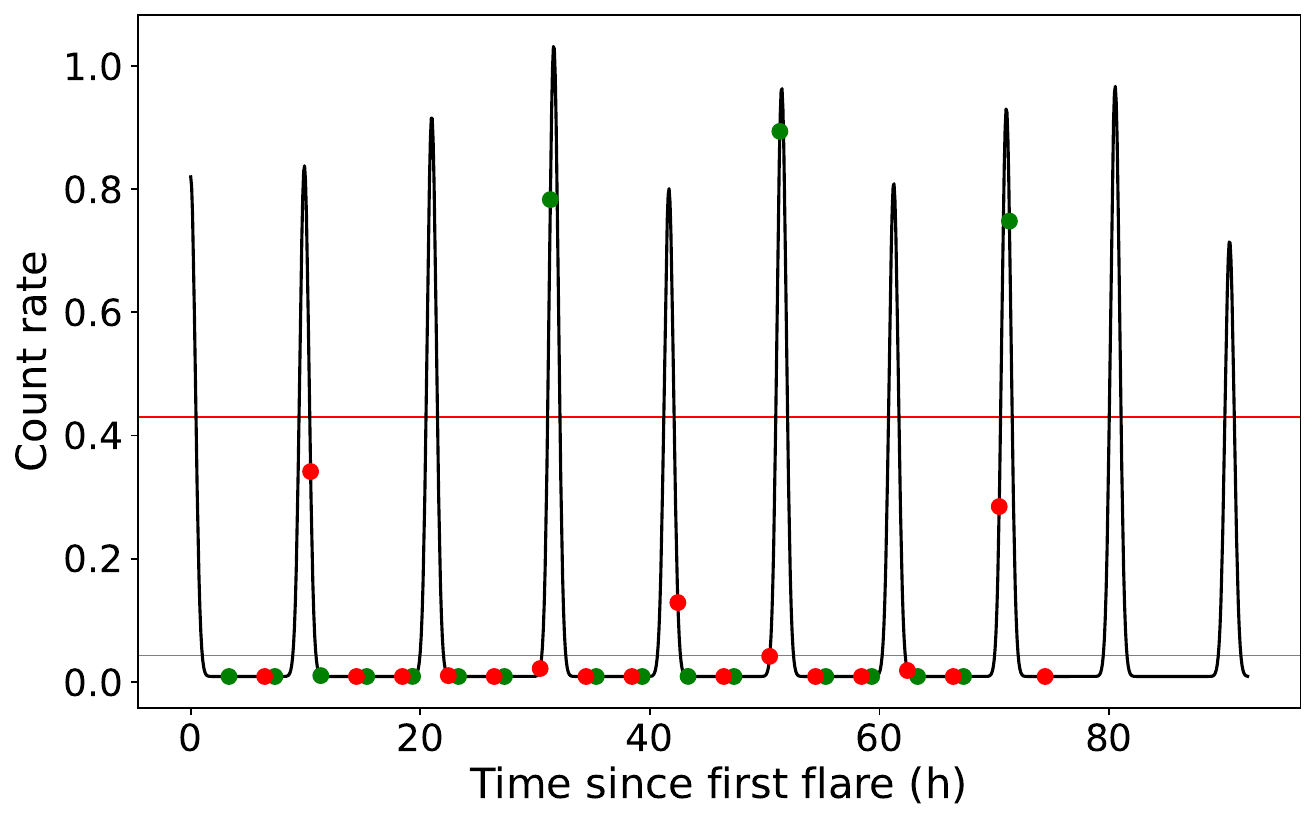}
		\caption{Example of simulated light curve, drawing from $\langle \log L_{\rm 0.5-2.0\,keV}^{\rm peak}\rangle \sim 42$ and $\langle t_{\rm recur}\rangle \sim 10\,$h within their $10\%$ scatter, and $\langle t_{\rm dur}\rangle \sim 1\,$h (from the relation in the bottom panel of Fig.~\ref{fig:time_fit}). An example of successful (\emph{green} points) and unsuccessful (\emph{red}) mock eROSITA scan is shown. The red line shows the eRASS sensitivity for bright states, the grey shows the threshold for faint states used in the simulations.}
		\label{fig:lc_simulated}
\end{figure}

We build 100 light curves for each grid value of $\langle \log L_{\rm 0.5-2.0\,keV}^{\rm peak}\rangle$ and $\langle t_{\rm recur}\rangle$ (and the inferred $\langle t_{\rm dur}\rangle$) and for two redshift bins ($z=0.02$ and 0.05, representative of the values spanned by known eROSITA QPEs, Fig.~\ref{fig:rates}). From $\langle \log L_{\rm 0.5-2.0\,keV}^{\rm peak}\rangle$ we obtain an average peak count rate using \texttt{WebPIMMS}, which we convert to a quiescent state assuming an amplitude of 100. Since the sensitivity within the single visit is typically much shallower than this value, the assumed amplitude is generally unimportant as the faint state will be eROSITA's background. On top of this quiescence, we generate Gaussian eruptions by drawing $\langle \log L_{\rm 0.5-2.0\,keV}^{\rm peak}\rangle$ and $\langle t_{\rm recur}\rangle$ within their $10\%$ scatter, with FWHM converted from 1/3 of the rise-to-decay duration inferred from the bottom panel of Fig.~\ref{fig:time_fit}. An example of generated light curve is shown in Fig.~\ref{fig:lc_simulated}, for $\langle \log L_{\rm 0.5-2.0\,keV}^{\rm peak}\rangle \sim 42$ and $\langle t_{\rm recur}\rangle \sim 10\,$h within their $10\%$ scatter, and $\langle t_{\rm dur}\rangle \sim 1\,$h (from Fig.~\ref{fig:time_fit}).

Each light curve has an associated efficiency value $\xi$, which can be seen as the probability that a given light curve would be seen by eROSITA, given its sampling, as QPE-like. This probability is computed as the fraction of successful scans out of the total possible, which are performed shifting a mock eRASS sampling through the light curve for enough times to cover a full QPE cycle. A scan is considered successful if at least two visits are found to be in a bright state, with at least a faint state in between, and combinations thereof (e.g., see the green points in Fig.~\ref{fig:lc_simulated}). A scan is otherwise considered unsuccessful (e.g., red points in Fig.~\ref{fig:lc_simulated}). A visit is considered in a high state if its count rate is larger than the sensitivity value of $\sim0.42\,$c/s adopted in Sec.~\ref{sec:erosita}, or in a low state if lower than a tenth of this value. For the brightest luminosity bins in the simulation, the quiescence state might be brighter than both these values, so we adopt a low state value as 1/50 of the peak and a high state value as 10 times the low. This is in qualitative analogy with the real QPE search presented in Sec.~\ref{sec:erosita}. We note that the choice of these threshold numbers have some impact on the single efficiency draw, but very little impact on the distribution after the 100 draws. Regarding the length of the mock eRASS baseline, we define a mock eRASS sampling as 18 data points, each with exactly $40\,$s exposure and separated by 4\,h (in comparison with the real analysis in Sec.~\ref{sec:erosita}). The latter is the typical revolution time of the field of view around the Ecliptic plane before coming back to the same point in the sky \citep{Merloni+2024:erass}. Instead, the number of visits simulated corresponds to two eRASS (i.e. 9 per eRASS) and it is based on the typical length of the baseline for the four eROSITA QPEs \citepalias{Arcodia+2021:eroqpes,Arcodia+2024:qpe34}. This preference for $\sim10-$eROday baselines is likely due to a sweet spot between having a large enough area in the sky and a long enough baseline. Longer baselines than these are obtained at the Ecliptic poles where, however, the area is small \citep[][]{Merloni+2024:erass,Bogensberger+2024:sep}. Since we correct for the efficiency given by the sampling, we do not expect that assuming this cadence imprints any significant bias on our calculations. Future work will be done in simulating a QPE population on the full sky and running mock source detection and light curve generation depending on the sky location, but it is beyond the scope of this work.

\begin{figure}[tb]
		\centering
		\includegraphics[width=0.95\columnwidth]{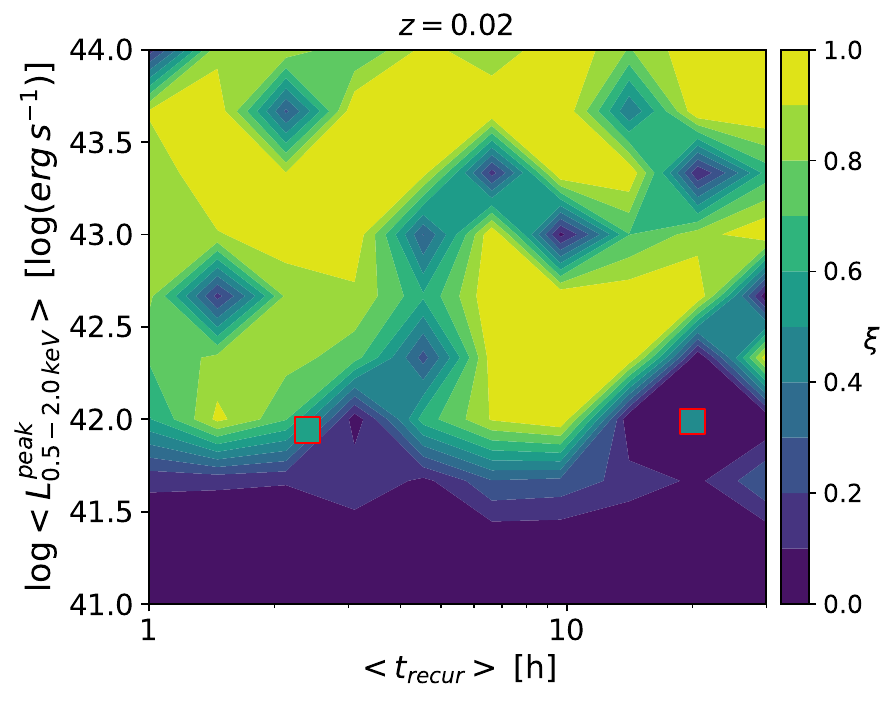}
		\includegraphics[width=0.95\columnwidth]{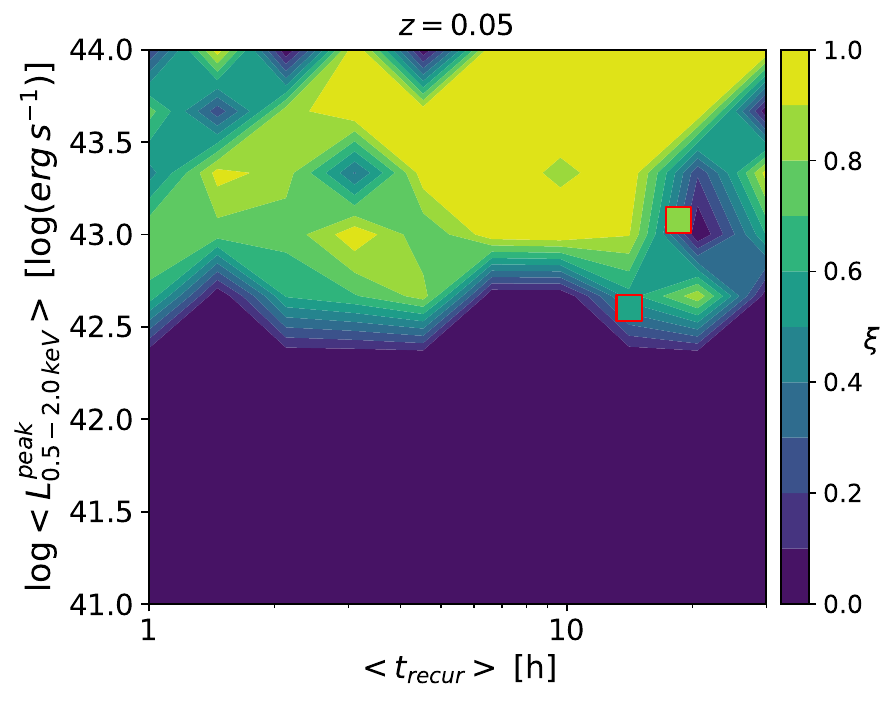}
	    \caption{Maps of the simulated eROSITA detection efficiency $\xi$ for QPE sources in a 2D grid of average peak luminosity ($\langle \log L_{\rm 0.5-2.0\,keV}^{\rm peak}\rangle$) and average recurrence time ($\langle t_{\rm recur}\rangle$), for the case of $z=0.02$ (\emph{top}) and $z=0.05$ (\emph{bottom}). Squares with a red contour highlight the known QPE sources in the respective redshift bins. Apart from grid points around periods resonating with eROSITA's 4\,h cadence, $\xi$ is fairly homogeneous above the cut provided by the sensitivity of each redshift bin.}
		\label{fig:simul_grids}
\end{figure}

\begin{figure}[tb]
		\centering
		\includegraphics[width=0.95\columnwidth]{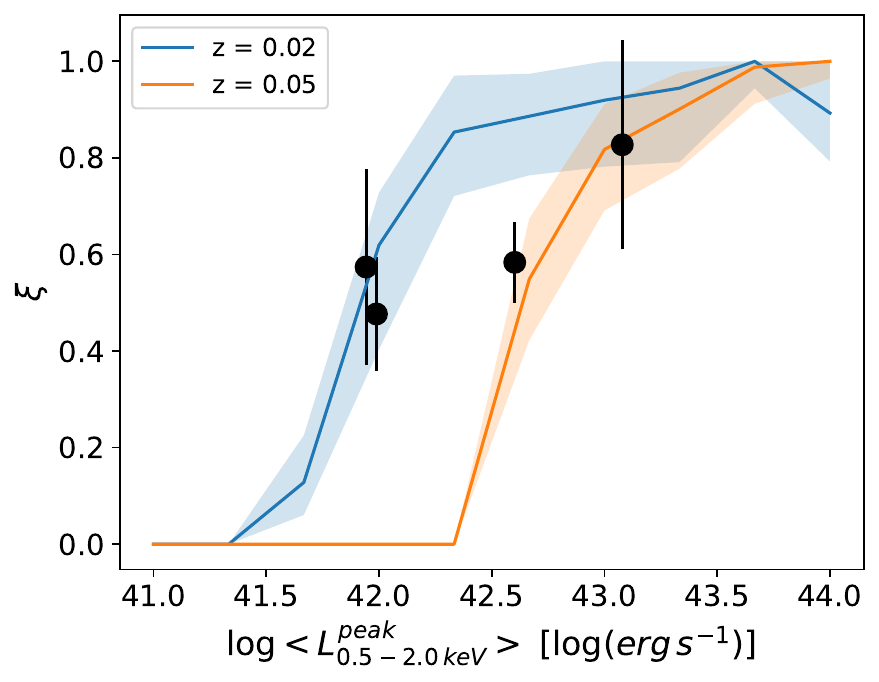}
	    \caption{Simulated detection efficiency $\xi$ as a function of peak luminosity $\langle \log L_{\rm 0.5-2.0\,keV}^{\rm peak}\rangle$ in the redshift bin $z=0.02$ (blue) and $z=0.05$ (orange). The plot is a version of Fig.~\ref{fig:simul_grids} collapsed on the y-axis. Black points indicate the known eROSITA QPEs \citepalias{Arcodia+2021:eroqpes,Arcodia+2024:qpe34}. They are representative of the $\xi$ values of the simulated population at their respective luminosity.}
		\label{fig:simul_eff_lpeak}
\end{figure}

After this procedure each light curve has an associated $\xi$, and each bin in the simulations has 100 light curves generated, thus 100 $\xi$ values. We show the outcome of these simulations in the two panels of Fig.~\ref{fig:simul_grids}, one per redshift bin adopted ($z=0.02$ and 0.05, respectively). The average $\xi$ per bin is mapped in a 2D grid of luminosity and recurrence. The first take-away is that in each redshift bin $\xi$ is fairly homogeneous above the sensitivity and that this sensitivity does not significantly depend on the timing properties ($\langle t_{\rm recur}\rangle$, $\langle t_{\rm dur}\rangle$). There are some obvious exceptions, corresponding to periods resonating with the $4$\,h cadence in the eRASS sampling. Further insights on the fact that eROSITA discoveries are likely not severely biased come from a comparison with their individual efficiencies (shown in Fig.~\ref{fig:hist}) compared to the simulated ones. We plot the known eROSITA QPEs as squares in Fig.~\ref{fig:simul_grids}, color coded in the same way. Their location is not in the regions were the efficiency is the highest, showing that the eROSITA discoveries have not selected the easiest QPEs to be found. An obvious exception is eRO-QPE3 which appears as an outlier in the top panel, surrounded by a region with $\xi\sim0$. This is probably due to the fact that eRO-QPE3 showed time-varying amplitude over the eRASS surveys \citepalias{Arcodia+2024:qpe34}, so our simplistic simulations have not grasped that sources can move across the grid. We also note that the average recurrence is only estimated crudely for eRO-QPE3 \citepalias{Arcodia+2024:qpe34}. Fig.~\ref{fig:simul_eff_lpeak} shows the 2D maps collapsed on the luminosity axis, for both redshift bins. It shows, even more quantitatively, that the QPEs discovered by eROSITA (shown as black points) are representative of the efficiency values within their luminosity bins. We conclude the known QPEs are, given our current knowledge on QPEs, a fair draw of their intrinsic population. Hence, this justifies their use to infer the volumetric rates presented in this work.

\section{Acknowledgments}

We thank the anonymous referee for their report, which improved the look of this manuscript. R.A. is grateful to I. Linial for the calculations on the stellar- versus BH-EMRI rates, and to A. Sesana, B. Metzger and N. Stone for insightful discussions on the volumetric rates and to E. Kara and R. Mushotzky for useful discussions on future X-ray missions. R.A. received support for this work by NASA through the NASA Einstein Fellowship grant No HF2-51499 awarded oby the Space Telescope Science Institute, which is operated by the Association of Universities for Research in Astronomy, Inc., for NASA, under contract NAS5-26555. GP acknowledges funding from the European Research Council (ERC) under the European Union’s Horizon 2020 research and innovation programme (grant agreement No 865637), support from Bando per il Finanziamento della Ricerca Fondamentale 2022 dell’Istituto Nazionale di Astrofisica (INAF): GO Large program and from the Framework per l’Attrazione e il Rafforzamento delle Eccellenze (FARE) per la ricerca in Italia (R20L5S39T9). We acknowledge the use of the matplotlib package \citep{Hunter2007:matplotlib}.

This work is based on data from eROSITA, the soft X-ray instrument aboard SRG, a joint Russian-German science mission supported by the Russian Space Agency (Roskosmos), in the interests of the Russian Academy of Sciences represented by its Space Research Institute (IKI), and the Deutsches Zentrum für Luft- und Raumfahrt (DLR). The SRG spacecraft was built by Lavochkin Association (NPOL) and its subcontractors, and is operated by NPOL with support from the Max Planck Institute for Extraterrestrial Physics (MPE). The development and construction of the eROSITA X-ray instrument was led by MPE, with contributions from the Dr. Karl Remeis Observatory Bamberg \& ECAP (FAU Erlangen-Nuernberg), the University of Hamburg Observatory, the Leibniz Institute for Astrophysics Potsdam (AIP), and the Institute for Astronomy and Astrophysics of the University of Tuebingen, with the support of DLR and the Max Planck Society. The Argelander Institute for Astronomy of the University of Bonn and the Ludwig Maximilians Universitaet Munich also participated in the science preparation for eROSITA. The eROSITA data shown here were processed using the eSASS software system developed by the German eROSITA consortium.

This work has made use of data from the European Space Agency (ESA) mission
{\it Gaia} (\url{https://www.cosmos.esa.int/gaia}), processed by the {\it Gaia}
Data Processing and Analysis Consortium (DPAC,
\url{https://www.cosmos.esa.int/web/gaia/dpac/consortium}). Funding for the DPAC
has been provided by national institutions, in particular the institutions participating in the {\it Gaia} Multilateral Agreement. This research has made use of the VizieR catalogue access tool, CDS, Strasbourg, France (DOI : 10.26093/cds/vizier). The original description of the VizieR service was published in 2000, A\&AS 143, 23.

The Legacy Surveys consist of three individual and complementary projects: the Dark Energy Camera Legacy Survey (DECaLS; Proposal ID \#2014B-0404; PIs: David Schlegel and Arjun Dey), the Beijing-Arizona Sky Survey (BASS; NOAO Prop. ID \#2015A-0801; PIs: Zhou Xu and Xiaohui Fan), and the Mayall z-band Legacy Survey (MzLS; Prop. ID \#2016A-0453; PI: Arjun Dey). DECaLS, BASS and MzLS together include data obtained, respectively, at the Blanco telescope, Cerro Tololo Inter-American Observatory, NSF’s NOIRLab; the Bok telescope, Steward Observatory, University of Arizona; and the Mayall telescope, Kitt Peak National Observatory, NOIRLab. Pipeline processing and analyses of the data were supported by NOIRLab and the Lawrence Berkeley National Laboratory (LBNL). The Legacy Surveys project is honored to be permitted to conduct astronomical research on Iolkam Du’ag (Kitt Peak), a mountain with particular significance to the Tohono O’odham Nation.

NOIRLab is operated by the Association of Universities for Research in Astronomy (AURA) under a cooperative agreement with the National Science Foundation. LBNL is managed by the Regents of the University of California under contract to the U.S. Department of Energy.

This project used data obtained with the Dark Energy Camera (DECam), which was constructed by the Dark Energy Survey (DES) collaboration. Funding for the DES Projects has been provided by the U.S. Department of Energy, the U.S. National Science Foundation, the Ministry of Science and Education of Spain, the Science and Technology Facilities Council of the United Kingdom, the Higher Education Funding Council for England, the National Center for Supercomputing Applications at the University of Illinois at Urbana-Champaign, the Kavli Institute of Cosmological Physics at the University of Chicago, Center for Cosmology and Astro-Particle Physics at the Ohio State University, the Mitchell Institute for Fundamental Physics and Astronomy at Texas A\&M University, Financiadora de Estudos e Projetos, Fundacao Carlos Chagas Filho de Amparo, Financiadora de Estudos e Projetos, Fundacao Carlos Chagas Filho de Amparo a Pesquisa do Estado do Rio de Janeiro, Conselho Nacional de Desenvolvimento Cientifico e Tecnologico and the Ministerio da Ciencia, Tecnologia e Inovacao, the Deutsche Forschungsgemeinschaft and the Collaborating Institutions in the Dark Energy Survey. The Collaborating Institutions are Argonne National Laboratory, the University of California at Santa Cruz, the University of Cambridge, Centro de Investigaciones Energeticas, Medioambientales y Tecnologicas-Madrid, the University of Chicago, University College London, the DES-Brazil Consortium, the University of Edinburgh, the Eidgenossische Technische Hochschule (ETH) Zurich, Fermi National Accelerator Laboratory, the University of Illinois at Urbana-Champaign, the Institut de Ciencies de l’Espai (IEEC/CSIC), the Institut de Fisica d’Altes Energies, Lawrence Berkeley National Laboratory, the Ludwig Maximilians Universitat Munchen and the associated Excellence Cluster Universe, the University of Michigan, NSF’s NOIRLab, the University of Nottingham, the Ohio State University, the University of Pennsylvania, the University of Portsmouth, SLAC National Accelerator Laboratory, Stanford University, the University of Sussex, and Texas A\&M University.

BASS is a key project of the Telescope Access Program (TAP), which has been funded by the National Astronomical Observatories of China, the Chinese Academy of Sciences (the Strategic Priority Research Program “The Emergence of Cosmological Structures” Grant \#XDB09000000), and the Special Fund for Astronomy from the Ministry of Finance. The BASS is also supported by the External Cooperation Program of Chinese Academy of Sciences (Grant \#114A11KYSB20160057), and Chinese National Natural Science Foundation (Grant \#12120101003, \#11433005).

The Legacy Survey team makes use of data products from the Near-Earth Object Wide-field Infrared Survey Explorer (NEOWISE), which is a project of the Jet Propulsion Laboratory/California Institute of Technology. NEOWISE is funded by the National Aeronautics and Space Administration.

The Legacy Surveys imaging of the DESI footprint is supported by the Director, Office of Science, Office of High Energy Physics of the U.S. Department of Energy under Contract No. DE-AC02-05CH1123, by the National Energy Research Scientific Computing Center, a DOE Office of Science User Facility under the same contract; and by the U.S. National Science Foundation, Division of Astronomical Sciences under Contract No. AST-0950945 to NOAO.

\end{appendix}
	
\end{document}